\documentclass[aps,prl,letterpaper,twocolumn,preprintnumbers,floatfix,superscriptaddress]{revtex4}
\usepackage{graphicx, epsfig}
\usepackage[dvipsnames]{xcolor}
\usepackage{hyperref}
\usepackage{mathrsfs}
\usepackage{amsmath}
\usepackage{enumerate}
\usepackage{bm}
\newcommand{\bmf}[1]{{\bm{#1}}}
\usepackage{float}
\usepackage{ulem}

\newcommand{\be}{\begin{equation}}
	\newcommand{\ee}{\end{equation}}
\newcommand{\bea}{\begin{eqnarray}}
	\newcommand{\eea}{\end{eqnarray}}
\newcommand{\ba}{\begin{eqnarray}}
	\newcommand{\ea}{\end{eqnarray}}

\newcommand{\gapp}{\mathrel{\raise.3ex\hbox{$>$}\mkern-14mu
		\lower0.6ex\hbox{$\sim$}}}
\newcommand{\lapp}{\mathrel{\raise.3ex\hbox{$<$}\mkern-14mu
		\lower0.6ex\hbox{$\sim$}}}

\begin{document}
	\title{Probing the $B+L$ violation process with the observation of cosmic magnetic field}
	
	\author{Yuefeng Di}
	
	\author{Jialong Wang}
	\affiliation{
		Department of Physics, Chongqing University, Chongqing 401331, China}
	\author{Ligong Bian }\email{lgbycl@cqu.edu.cn}
	\affiliation{
		Department of Physics, Chongqing University, Chongqing 401331, China}
	\affiliation{
		Chongqing Key Laboratory for Strongly Coupled Physics, Chongqing 401331, China}
	
	\author{Rong-Gen Cai}\email{cairg@itp.ac.cn}
	\affiliation{CAS Key Laboratory of Theoretical Physics, Institute of Theoretical Physics, Chinese Academy of Sciences, Beijing 100190, China}
	\affiliation{School of Physical Sciences, University of Chinese Academy of Sciences, Beijing 100049, China}
	\affiliation{School of Fundamental Physics and Mathematical Sciences, Hangzhou Institute for Advanced Study, University of Chinese Academy of Sciences, Hangzhou 310024, China}
	\author{Jing Liu}\email{liujing@ucas.ac.cn}
	\affiliation{School of Fundamental Physics and Mathematical Sciences, Hangzhou Institute for Advanced Study, University of Chinese Academy of Sciences, Hangzhou 310024, China}
	\affiliation{School of Physical Sciences, University of Chinese
		Academy of Sciences, Beijing 100049, China}

	\begin{abstract}
		
		We investigate the $B+L$ violation process by performing three-dimensional lattice simulations in an electroweak theory with first-order phase transition and the electroweak sphaleron decay. The simulation results indicate that the Chern-Simons number changes along with the helical magnetic field production when the sphaleron decay occurs. Our study suggests that, for the electroweak phase transition with nucleation rate being smaller than $\sim \mathcal{O}(10)$, the helical magnetic field with the fractional magnetic helicity $\epsilon_M\leq 0.2$ can be probed by Cherenkov Telescope Array through the intergalactic magnetic field measurements. Based on our numerical results, we suggest a  method to probe the baryon asymmetry generation of the Universe, which is a general consequence of the electroweak sphaleron process, through the astronomical observation of the corresponding helical magnetic field.

	\end{abstract}
	
	\maketitle
	
	\noindent{\it \bfseries  Introduction:}
	While the phase transition in the Standard Model (SM) is {\it cross-over}~\cite{DOnofrio:2014rug}, the detection of the first-order phase transition (PT) generally predicted in many new physics models would indicate the new physics discovery beyond the SM. Besides the generation of the stochastic gravitational waves (GWs)~\cite{Mazumdar:2018dfl,Caprini:2015zlo,Caprini:2019egz} in a first-order phase transition, it was also proposed to generate magnetic fields (MFs) which may seed  cosmological MFs~\cite{Durrer:2013pga,Grasso:2000wj,Subramanian:2015lua,Kandus:2010nw} observed by extensively astronomical observations~\cite{Yamazaki:2012pg,2019MNRAS.486.4275X,2018ApJS..234...11H,doi:10.1146/annurev-astro-091916-055221}. The mechanism to generate MFs during the PT come from two aspects, bubble collisions~\cite{Zhang:2019vsb,Di:2020ivg} and the Chern-Simons (CS) number-violating electroweak sphaleron decay~\cite{Vachaspati:2001nb,Copi:2008he,Cornwall:1997ms,Mou:2017zwe}, which have been investigated separately. The latter gives rise to the specific interest for its ability to generate helical MFs and validate the $B+L$ violation process which further connects with the generation of the baryon asymmetry in the universe~\cite{Kuzmin:1985mm,Rubakov:1996vz,Vachaspati:2001nb,Cornwall:1997ms}. 
	Since the electroweak sphaleron process in the broken phase is exponentially suppressed after the completion of the PT \footnote{In the well-motivated electroweak baryogenesis paradigm, it was argued that the first-order PT is intensively connected with the electroweak sphaleron process in electroweak theories, which requires that the sphaleron rate in the broken phase (inside the electroweak bubbles) should be smaller than the Hubble rate during the PT~\cite{Morrissey:2012db}. }, it is hard to detect at high energy colliders~\cite{Rubakov:1996vz}\footnote{However, see Ref.~\cite{Ellis:2016dgb} for a different opinion.}, the observation of helical MFs in galaxy clusters provides a new probability to explore the electroweak sphaleron process and the baryon asymmetry.  
	
	The CS number evolution during the first-order PT process has not been studied in the literature. In this Letter, we numerically investigate the complete process of a first-order electroweak PT with the decay of the electroweak sphaleron, including the CS number evolution and the MF production\footnote{For the study of the generation of MFs during preheating within a scenario of hybrid inflation utilizing electroweak model we refer to Ref.\cite{PhysRevLett.100.241301,D_az_Gil_2008}.}. We adopt the field configurations for the electroweak sphaleron inside the true vacuum bubbles to study the effect of the sphaleron decay process, then obtain the relationship between the CS number evolution and the electroweak sphaleron decay, and figure out the generated MF spectrum. Our work demonstrates the possibility to probe the electroweak sphaleron decay through the MF observations. 
	We choose $c=1$ throughout this Letter and neglect the expansion of the Universe since we consider the duration of the first-order PT is much shorter than the Hubble time.

	\noindent{\it \bfseries  PT with sphaleron decay: }
	\label{sec:model}
	The relevant bosonic part of the Lagrangian in the electroweak theory is,
	\begin{equation}
		\mathcal{L}=\vert D_\mu \Phi\vert^2-\frac{1}{4} W^a_{\mu\nu} W^{a\mu\nu}
		- \frac{1}{4} B_{\mu\nu} B^{\mu\nu}-V(\Phi,T)\;.
		\label{lagrangian}
	\end{equation}
	Here, the covariant derivative is
	$
	D_\mu = \partial_\mu - i \frac{g}{2} \sigma^a W^a_\mu - i \frac{g'}{2} B_\mu$, where $\sigma^a$ ($a=1,2,3$) are the Pauli matrices,
	and $g=0.65$, $g'=0.53g$. The
	$\Phi$ and $V(\Phi,T)$ denote the Higgs field and the Higgs potential. The Higgs potential for a first-order PT requires going beyond 
	the SM, such as: SM extended with a dimensional-six operator $(\Phi^\dagger \Phi)^3/\Lambda^2$~\cite{Grojean:2004xa,Grojean:2006bp},  xSM~\cite{Profumo:2014opa,Zhou:2019uzq,Zhou:2020idp,Alves:2018jsw,Profumo:2007wc,Espinosa:2011ax,Jiang:2015cwa}, 2HDM~\cite{Cline:2011mm,Dorsch:2013wja,Dorsch:2014qja,Bernon:2017jgv,Andersen:2017ika,Kainulainen:2019kyp}, George-Macheck model~\cite{Zhou:2018zli}, and NMSSM~\cite{Bian:2017wfv,Huber:2015znp}. Here, the Higgs potential $V(\phi,T)$ embracing a barrier at finite temperature can trigger a first-order PT proceeding with bubble nucleations and collisions~\cite{Grojean:2004xa} that are expected to produce MFs
	~\cite{Durrer:2013pga,Kandus:2010nw,Subramanian:2015lua} and GWs~\cite{Grojean:2006bp}\footnote{ Model parameters that determining $V(\phi,T)$ at the PT temperature are related with bubble characteristics (mean bubble separation, wall thickness and wall velocity)~\cite{Hindmarsh:2013xza,Cutting:2019zws,Hindmarsh:2015qta,Hindmarsh:2017gnf}. }.
	We study MFs produced by sphaleron decay and bubble collisions.
	The initial conditions are set as
	$\Phi=\dot\Phi=0$ before the beginning of bubble nucleation. The  nucleated 
	bubble profile is  assumed as 
	\begin{equation}
		\label{eq:Phiini}
		\Phi (t=0,{\bm r}) =  \frac{v}{2} \left [ 1-\tanh \left ( \frac{r-R_0}{L_{\rm w}}\right ) \right ]
		\begin{pmatrix} 0 \\ 1 \end{pmatrix} \;,
	\end{equation}
	where $r=R_0$ is the initial nucleated bubble radius and $L_{\rm w}$ is set as the thickness of the critical bubble wall, which can be obtained through the potential barrier at the PT.  Inside the nucleated bubble, we consider Higgs and W field configurations like the electroweak sphaleron 
	solution~\cite{Manton:1983nd,Klinkhamer:1984di} 
	\be
	\Phi = \frac{v h(\xi)}{r_s} \begin{pmatrix}
		i x + y \\
		-i z
	\end{pmatrix}
	\ee
	\bea
	W_i^a {\bf \tau}^a = -\frac{2 f(\xi)}{g r_s^2} \epsilon_{i c b} x_b {\bf G}_\Theta  
	{\bf \tau}^c {\bf G}_\Theta^\dagger 
	\label{sphcon}
	\eea
	where
	\begin{displaymath}
		{\bf G}_\Theta \left(\vec{x}\right)=\exp{\left[i \Theta(r_s) 
			{\bm \tau} \cdot \hat{\bf x}/2 \right]}
	\end{displaymath}
	is an $SU(2)$ gauge transformation, with $\Theta(r_s)=\pi \left(1-\exp(-r_s)
	\right)$, ensuring that the gauge fields fall off fast enough away from
	the origin.
	The $f(\xi)$, $h(\xi)$ are profile functions with radial coordinate of $\xi=g v r_s/\sqrt{2}$. For this study, 
	we adopt  the profile functions of Klinkhamer and Manton's {\it Ansatz a}~\cite{Klinkhamer:1984di}, and
	the length scales are taken as $\Xi = 3.79$ and $\Omega = 1.90$, which ensures the decay of sphaleron. The radius of sphaleron configuration ($R_s$) should be smaller than the 
	radius of nucleated bubbles ($R_0$), and the magnitudes of the Higgs and gauge fields approach to theses values in the broken phase as the radius of the sphaleron configurations approaches to $R_s$ ($r_s\rightarrow R_s$). With this setup, the three scales of $R_s$, $R_0$, and $L_{\rm w}$ are connected through the relation $R_{\rm w}=R_0-R_s+L_{\rm w}/2$ with  $R_{\rm w}$ being the bubble regions between the bubble wall and $R_s$. 
	The effect of $U(1)$ gauge fields is neglected here. 
	Vacuum bubbles randomly nucleate in the
	regions where the symmetry is unbroken. We take the temporal gauge with $W_0^a = B_0=0$ and evolve equations of motion for relevant bosonic fields on the lattice as in Ref.~\cite{Di:2020ivg} to generate the MFs.

	\noindent{\it \bfseries CS number and MF production:}
	During the baryogenesis, the change of the baryon number is associated with the change of the Chern-Simons number formed by the $B+L$ violation through the anomaly equation of $\Delta N_B=N_F\Delta N_{CS}$ with $N_F =3$ being the number of fermion families. The formalism of $N_{CS}$ in our simulation is
	\begin{eqnarray}
		N_{CS} (t) &=& \frac{1}{32\pi^2} \epsilon^{ijk} \int d^3 x  \biggl [ 
		- {g'}^2 B_{ij} B_k  
		\nonumber \\
		&+& g^2 \left ( W_{ij}^a W_k^a - \frac{g}{3} 
		\epsilon_{abc} W_i^a W_j^b W_k^c \right )
		\biggr ]
		\label{NCSformula}
	\end{eqnarray}
	Adopting the sphaleron configuration given in Eq.~\ref{sphcon}, we numerically study the CS number evolution, associated with the sphaleron decay and the magnetic helicity at the PT which will be explored later. 
	
	For the study of the MFs production, we define the electromagnetic field after the Higgs field leaves the symmetric phase as
	$
	A_\mu = \sin \theta_w n^a W^a_\mu +  \cos \theta_w B_\mu$,
	where $\theta_w$ is the weak mixing angle satisfying $\sin^2\theta_w=0.22$, and $n^a \equiv -(\Phi^\dagger \sigma^a \Phi)/v^2$ presents the direction of the Higgs field.
	The corresponding field strength is constructed
	as~\cite{tHooft:1974kcl,Vachaspati:1991nm}
	\begin{equation}
		\begin{split}
			A_{\mu\nu} =& \sin \theta_w n^a W^a_{\mu\nu} + \cos \theta_w B_{\mu\nu} 
			- i\frac{2}{g v^2} \sin \theta_w\\ &\times \left[ (D_\mu\Phi)^\dagger (D_\nu \Phi) - (D_\nu\Phi)^\dagger (D_\mu \Phi) \right]\;.
			\label{eq:amunu}
		\end{split}
	\end{equation}
	We note that the field strength tensor (Eq.~\ref{eq:amunu}) for the calculation of MF strength includes the SU(2)$_L$ and U(1)$_Y$ field strengths $W^a_{\mu \nu}$ and $B_{\mu \nu}$, and the Higgs gradients part,
	depending on bubble collision dynamics and sphaleron configurations as will be seen later. 
	Following the conventions in Ref.~\cite{Brandenburg:2017neh, Brandenburg:2018ptt},  MFs
	can be described in terms of the equal-time correlation function of $\langle B_i^* (\bmf{k},t )B_j(\bmf{k}',t)\rangle
	= (2\pi )^3 \delta^{(3)}(\bmf{k}-\bmf{k}' )F_{ij} (\bmf{k},t)$,
	where $ B_i (\bmf{k},t )$ is the Fourier transformation of $ B_i (\bmf{x},t )$.
	Taking into account the helical (antisymmetric) part of $ F_{ij} (\bmf{k},t)$, one has
	$
	F_{ij} (\bmf{k},t)/(2\pi )^3
	= (\delta_{ij}-\hat{k}_{i}\hat{k}_{j}) E_{M}(k,t)/(4\pi k^{2})+i\epsilon_{ijl}k_l H_M/(8 \pi k^2)$, where $E_M$, the magnetic energy density can be obtained as~\cite{Durrer:2013pga}:
	$\rho_B \left(t\right) =\int_{0}^{\infty}E_{M}\left(k,t\right)dk$.
	Therefore, the MF strength can be obtained as:
	$
	B_\xi=\sqrt{2d\rho_B/d\log (k)}\;
	$.
	The observable root mean square (scale-averaged) MF strength, at the ``characteristic" correlation length
	being $
	\xi_M(t)=\int dk k^{-1} E_{M}\left(k,t\right)/\rho_B \left(t\right)$, can be estimated as
	~\cite{Brandenburg:2017neh} 
	$: B_{rms}(t)=\sqrt{2\rho_B \left(t\right)}$.
	The averaged helicity density of a magnetic field in a volume V is,
	\begin{eqnarray}
		\mathcal{H_M}=\frac{1}{V}\int d^3x {\bf A}\cdot {\bf B}\equiv\int dk H_M(k)\;,
	\end{eqnarray} 
	with ${\bf B}={\bm \nabla}\times {\bf A}$. 
	The magnetic energy spectrum $E_M(k)$ and the magnetic helicity spectrum are bounded by the realizability condition in the wavenumber space~\cite{Durrer:2003ja,Brandenburg:2004jv},
	\begin{eqnarray}
		|H_M(k)|\leq\frac{2 E_M(k)}{k}\;,
	\end{eqnarray}  
	with the equality condition met when the magnetic field reaches the maximally helical configuration. 
	The fractional magnetic helicity can be defined as~\cite{Brandenburg:2017neh}
	\begin{equation}
		\epsilon_M(t)
		=\frac{ \mathcal{H_M}(t)}{2 \xi_M(t) \rho_B(t)}\;,
		\label{sigma}
	\end{equation}
	which determines the evolution of the MFs generated from the PT after taking into account turbulent effects.
	
	\noindent{\it \bfseries Numerical results:\label{sec:Numerical-Simulation}}
	Our simulations are performed on a cubic lattice with the resolution $256^{3}$, and the lattice size $L^{3}$ is related with the number of bubbles initially placed in the lattice. The spatial resolution is taken as $\Delta x=L/256$, and the time spacing is chosen to be $\Delta t=L/1280$. We have verified that the lattice spacing chosen here is enough to ensure that we can capture all the dynamics for the generation of MFs.
	The mean bubble separation in the simulations is $R_\star=(L^3/N_b)^{1/3}$ with $N_b$ being the number of the generated bubbles at the simulation time, which determines the Lorentz factor for the bubble to be $\gamma_\star=R_\star/(2 R_0)$, and the radius of sphaleron configuration to be $R_s^\star=R_s/\gamma_\star$ at the bubble collision time. To illustrate the effect of the sphaleron decay on the MF production, we consider the two cases of nucleation of vacuum bubbles with and without considering sphaleron. One sphaleron configuration is implemented in each bubble when considering sphaleron. 
	For the case of $N_b=19$, we have $\gamma_\star=2.44$, with the wall velocity being $v_w=0.77$ and the mean bubble separation $R_\star\equiv 75.07 (L_{\rm w} /\gamma_\star)$ at the time of bubble collision. 
	
	\begin{figure}[!htp]
		\begin{center}
			\includegraphics[width=0.4\textwidth]{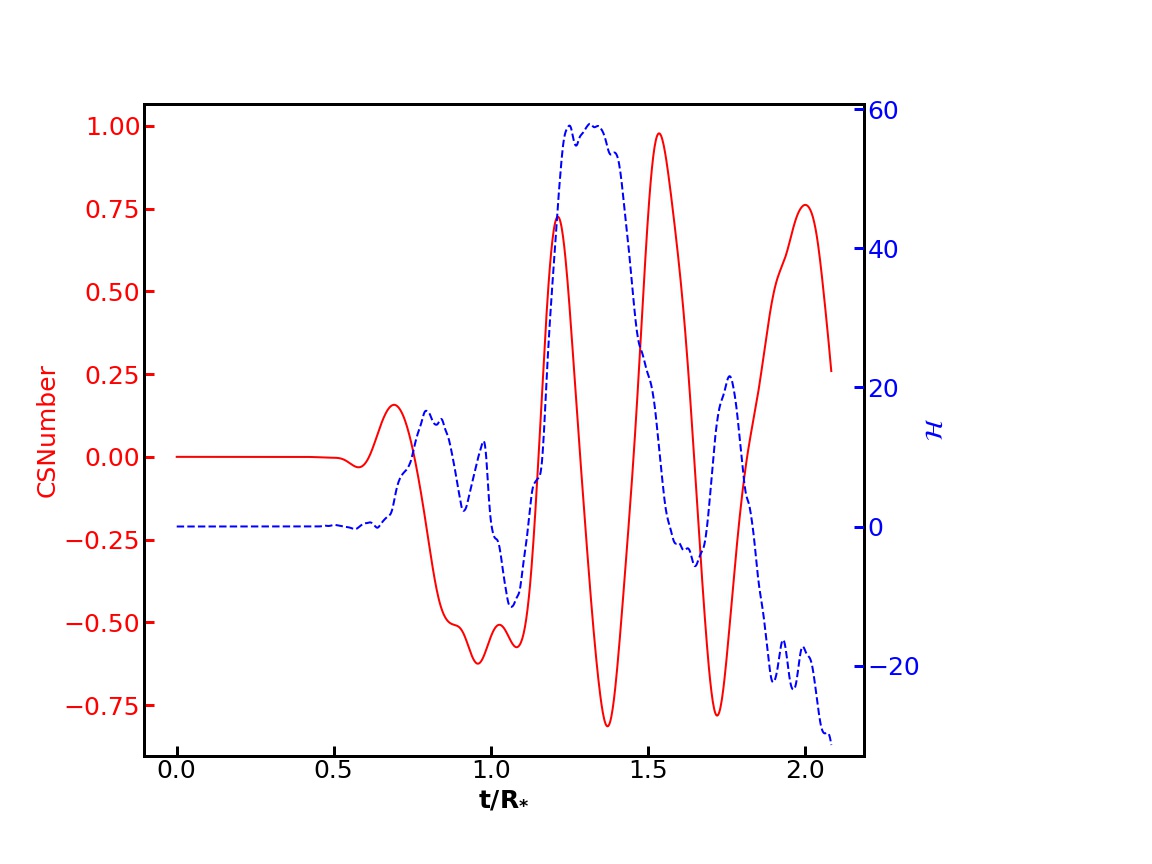}
			\includegraphics[width=0.4\textwidth]{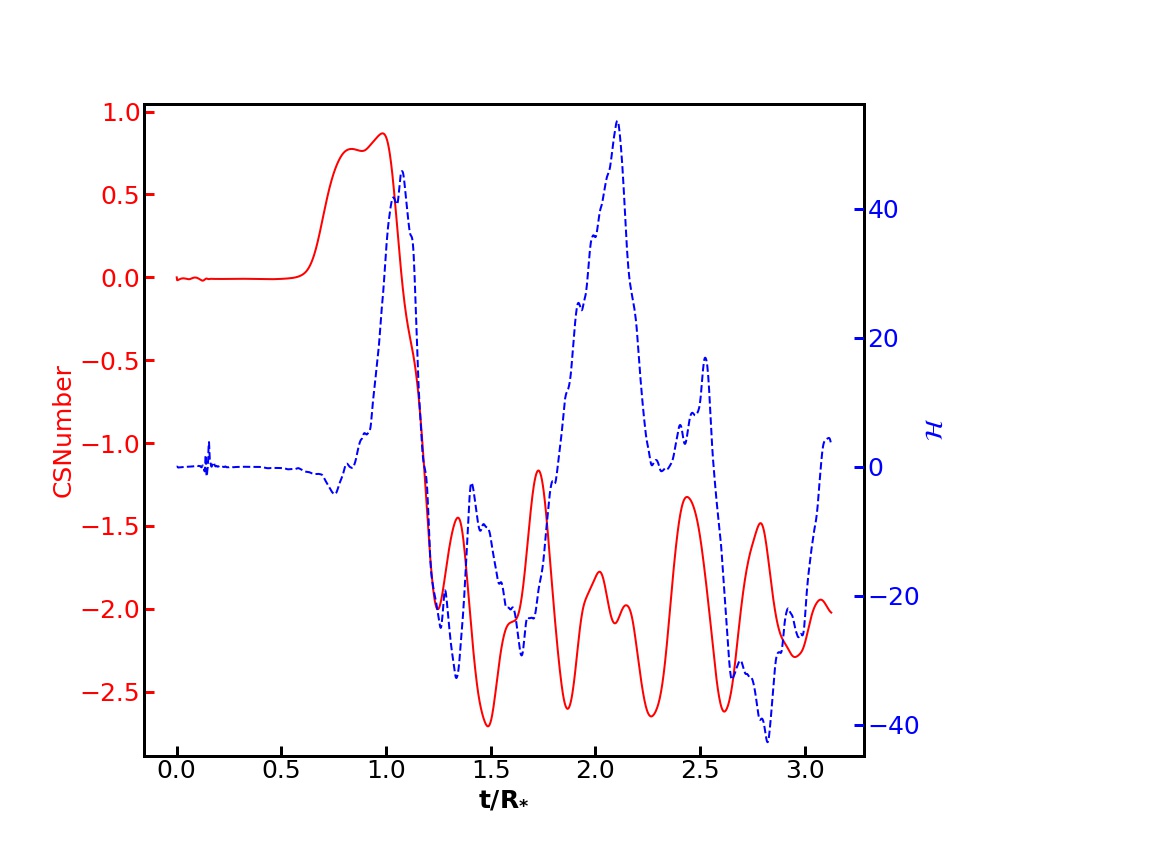}
			\caption{The Chern-Simons number of $3N_{CS}$ and the magnetic helicity for the scenario of $N_b=19$ without sphaleron (top), and for  the scenario with $R_s=12$ (bottom).   } \label{CS}
		\end{center}
	\end{figure}
	
	After the vacuum bubbles nucleated at the onset of the phase transition, the sphaleron configuration appears. The bubbles would expand due to the vacuum energy, and the PT completes with bubble coalescence after bubbles collide with each other. Meanwhile, the sphaleron decay occurs later with the sphaleron configurations disappear when the bubble collision occurs (with $t/R_\star \sim 1$), but it results in a non-zero helicity in produced MF. 
	In Fig.~\ref{CS}, we present the magnetic helicity and the Chern-Simons number evolution during the PT for the first time.  The oscillations of the magnetic helicity and the CS number during the evolution are observed, which is due to oscillations of the Higgs field and gauge fields associated with the bubble dynamics.     
	We find that : 1) for the scenario without the sphaleron configurations inside the electroweak bubbles, the
	CS number oscillates around zero throughout the PT process;
	2)  the CS number decreases from $1$ to $-2$ due to the sphaleron decay, the 
	change of the $\Delta N_{CS}=-1$ occurs around $t/R_\star\sim 1$, with the $B+L$ anomaly equation we obtain $\Delta N_B=3$ and the relationship between the magnetic helicity and $N_{CS}$ (and the baryon number $N_B$) is found to be $|\mathcal{H}|\sim 18 \Delta N_{CS} (\sim 6 \Delta N_B)$.
	
	\begin{figure}[!htp]
		\begin{center}
			\includegraphics[width=0.4\textwidth]{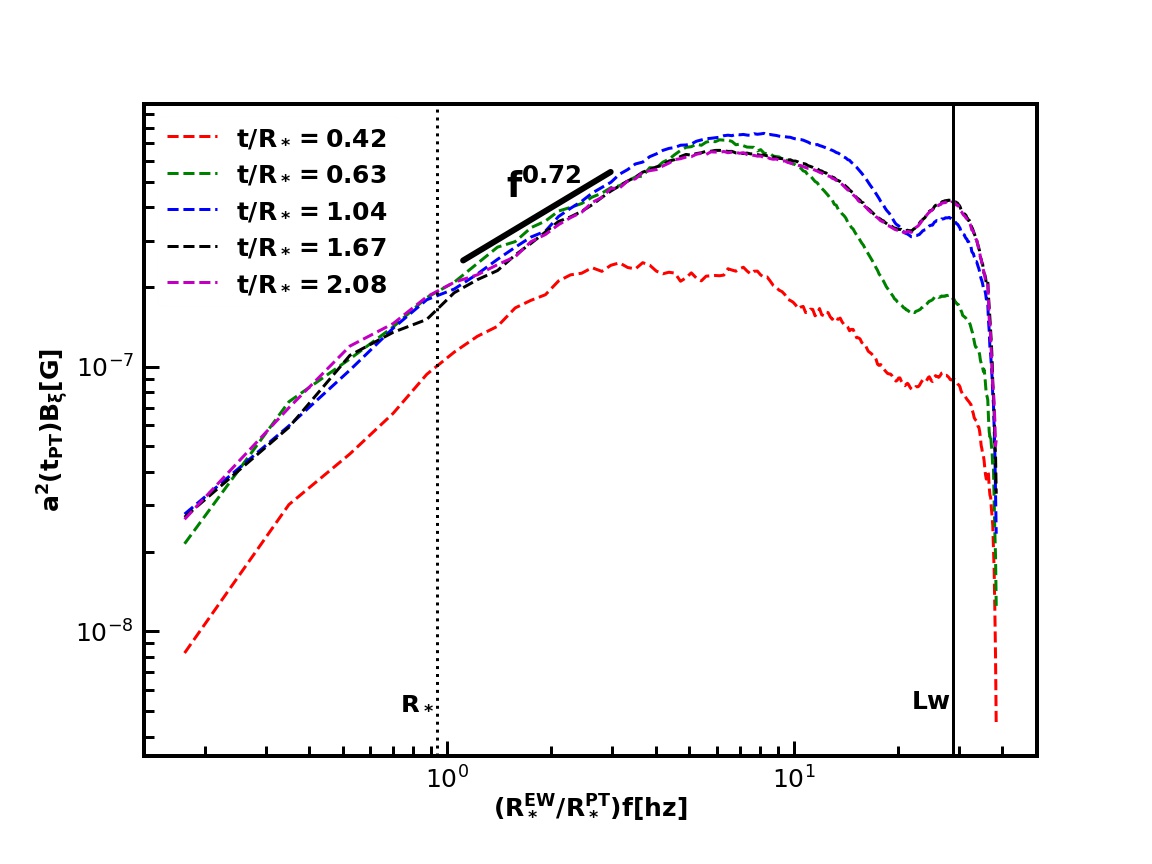}
			\includegraphics[width=0.4\textwidth]{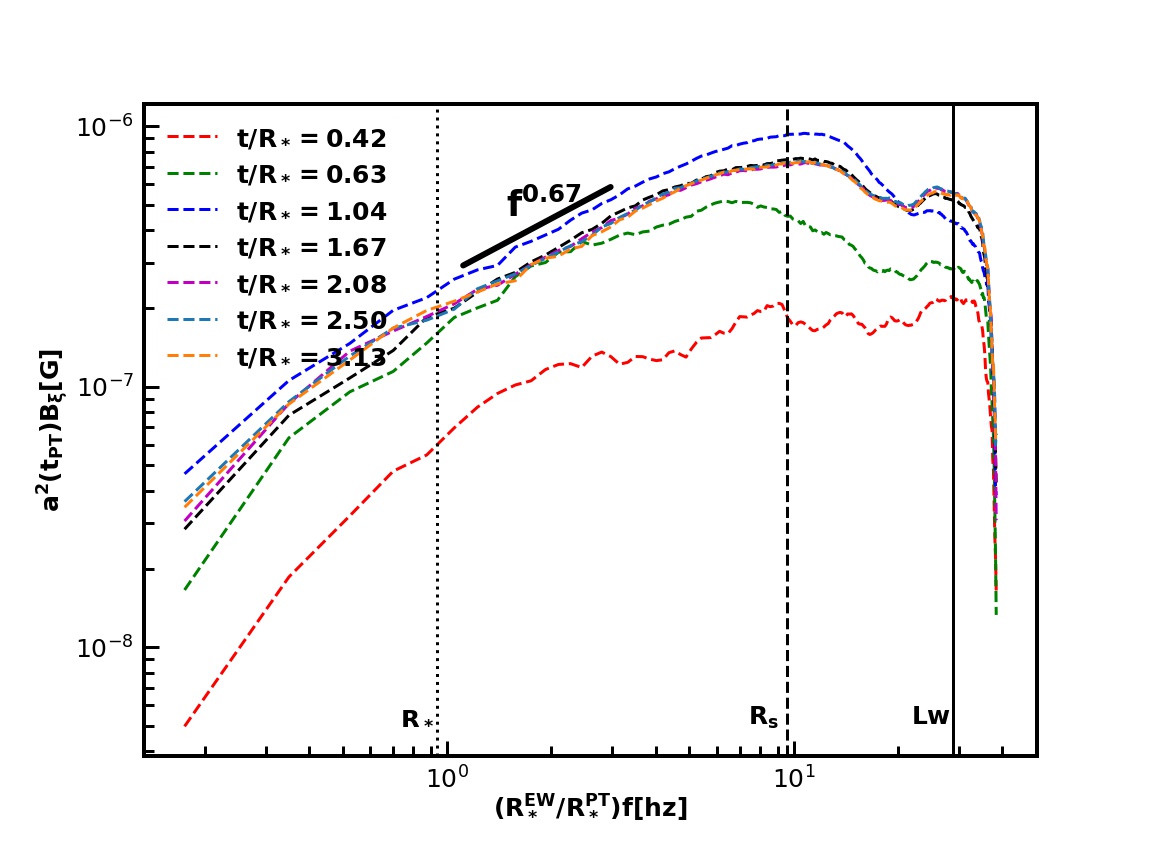}
			\caption{The frequency-dependent MF spectra during the first-order PT process with $N_b=19$ for the scenarios without the  sphaleron (top), and with sphaleron radius of $R_{s}=12$ (bottom). The length scales associated with $R_\star$, $R_{s}$, and $L_{\rm w}$ are plotted as vertical black dotted line, vertical dashed line, and vertical solid line. We perform the simulation until the times that are long enough for the generation of MFs, as the stop growth of amplitudes of MF can be found there.} \label{pmf}
		\end{center}
	\end{figure}
	
	We then numerically calculate the spectrum of the MF strength generated during the first-order electroweak PT (the PT temperature $T=100$ GeV is set here)~\cite{Di:2020ivg}. The sphaleron decay yields MF at the early time when the bubble nucleates before bubble collisions and coalescence, with  $R_s$ close the peak region of the MF spectrum at the generation times, as depicted in Fig.~\ref{pmf}.  
	The MF spectrum on the right hand side of the line of $L_{\rm w}$ is not reliable because the scale there is close to  $\Delta x$. Through the comparison of the two plots, we find that the sphaleron decay contributes to the high-frequency region of the MFs production. And, with the proceeding of the PT process, the magnitude of MFs from bubble collisions and Higgs oscillations (in the low-frequency region) would increase by around one order and therefore dominates over the MFs production from the sphaleron decay, as indicated by the figure. 
	To our knowledge, the behavior of the MFs spectrum from the sphaleron decay during bubble expansion and collision is observed for the first time, with the power-law of the MF spectrum in the infrared region being proportional to $f^{0.67}$ for the scenario with sphaleron radius $R_s=12$ (see bottom panel). 
	The power-law index in the low-frequency region is found to be slightly shallower than the spectrum generated without considering the sphaleron decay contribution (see the top panel, with $a^2 B_\xi\sim f^{0.72}$). The MF strength reaches $B_\xi \sim \mathcal{O}(10^{-7})$ G at the peak frequency for the two scenarios. The mean bubble separation parameter $R_\star$ is a function of bubble nucleation rate $\beta\sim \mathcal{O}(10-10^{3})H_\star$ for the electroweak first-order phase transition that can produce detectable GWs~\cite{Caprini:2019egz}. For this case, the correlation length of the MF at the generation time is estimated to $\xi_\star\sim 4.46\times 10^{-10}/ (\beta/H_\star)$ Mpc.

	\begin{figure}[!htp]
		\begin{center}
			\includegraphics[width=0.4\textwidth]{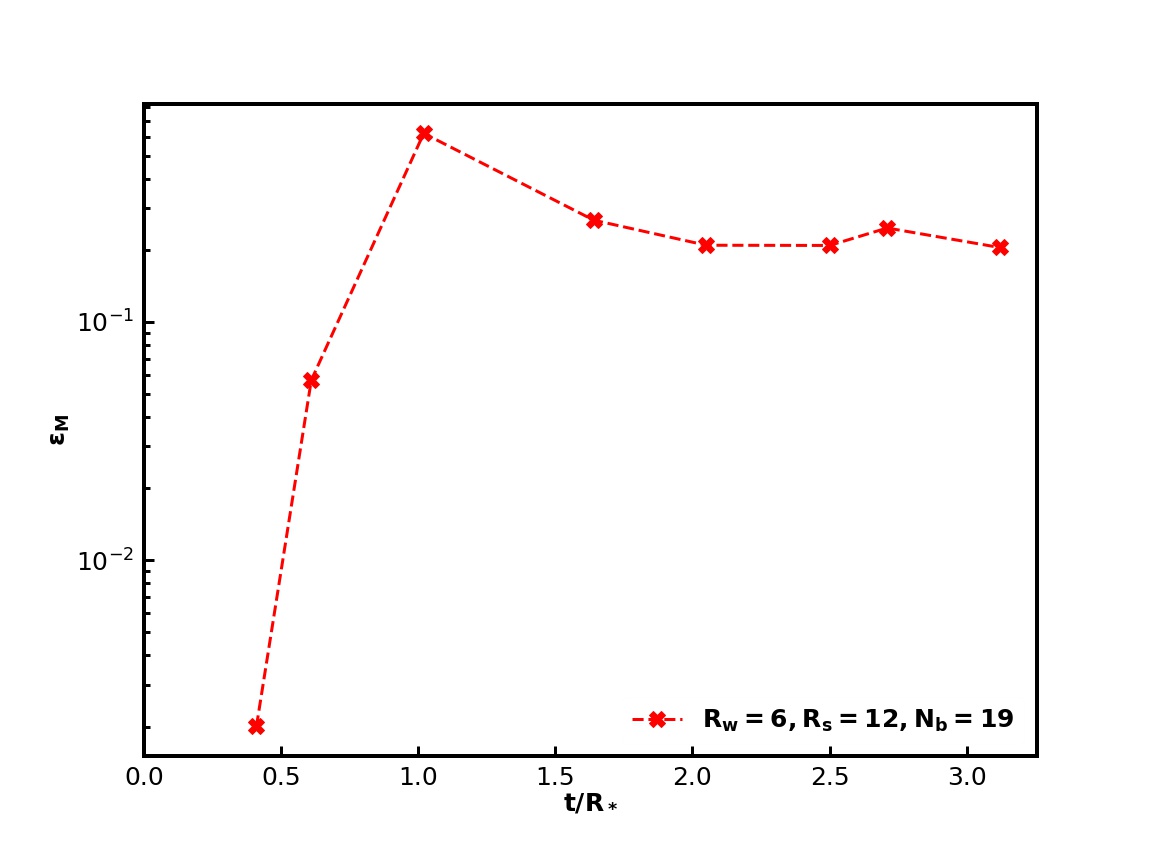}
			\includegraphics[width=0.4\textwidth]{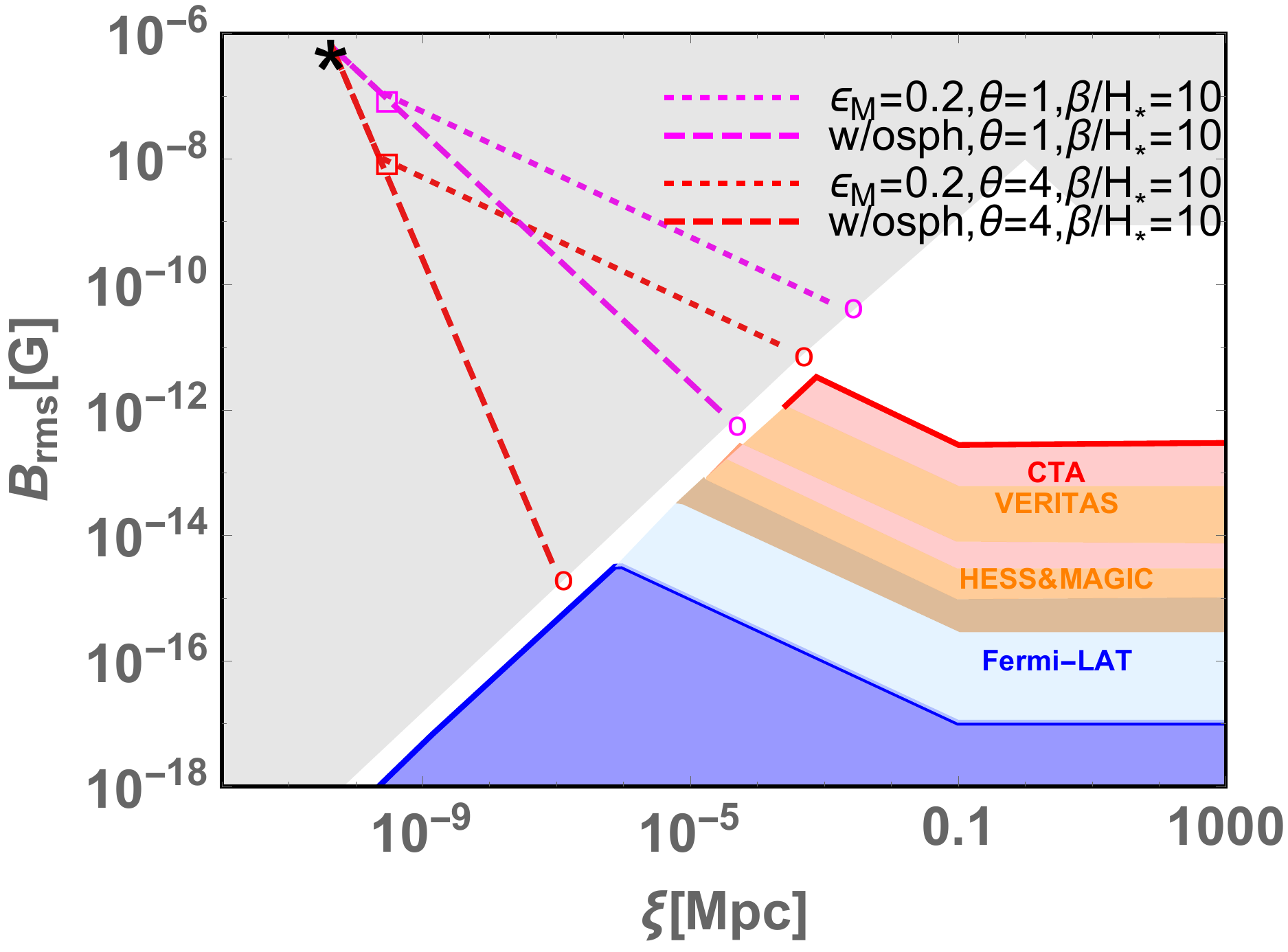}
			\caption{Top: The time dependence of the fractional magnetic helicity $\epsilon_M$.  Bottom: MFs evolution. In the first stage, MFs reach the maximum helicity (marked with squares), and finally evolve to the recombination time (marked with circles).  Gray region are excluded by MHD turbulent decay and CMB anisotropy measurements~\cite{Durrer:2013pga}. Projected sensitivity of the CTA observations to IGMF strength are marked with red region~\cite{CTA:2020hii}. Orange regions are excluded at the 95\% to 99\% confidence level from H.E.S.S., MAGIC~\cite{HESS:2014kkl}\cite{MAGIC:2010goh} and VERITAS~\cite{VERITAS:2017gkr}. The dark and light blue regions are excluded by the lower bounds on IGMF strength from Fermi-LAT~\cite{Ackermann_2018}\cite{Taylor:2011bn}. } \label{helicity}
		\end{center}
	\end{figure}

	We evaluate the fractional magnetic helicity $\epsilon_M$ for the scenario with $R_s=12$, which shows that $\epsilon_M$ grows to be a constant after the time of $t/R_\star>1$, see Fig.~\ref{helicity}. 
	Adopting $\epsilon_M$ at the end of the simulation, we calculate the root mean squared magnetic filed at correlation length after taking into account the turbulent evolution of MFs generated in the PT as simulated by Ref.~\cite{Brandenburg:2017neh}. For the fractional magnetic helicity, the magnetic field reaches the 
	maximum helicity at the first stage according to the scaling law $\xi_{hel}=\xi_\star(t_{hel}/t_\star)^{q_\theta}, B_{hel}=B_\star(t_{hel}/t_\star)^{-p_\theta/2}$ with  $t_{hel}=t_\star(\epsilon_{M,\star})^{-1/q_\theta}$ and $q_\theta=2/(\theta+3), p_\theta=2(\theta+1)/(\theta+3)$, 
	and then evolve to the recombination time $t_{rec}$ following $\xi_{rec}=\xi_{hel}(t_{rec}/t_{hel})^{q_0},B_{rec}=B_{hel}(t_{rec}/t_{hel})^{-p_0/2}$. We take $\theta=1,4$ and $p_0=q_0=2/3$ to evaluate the magnetic field strength and the correlation length at the recombination time, with the results shown in the bottom panel of Fig.~\ref{helicity}. In contrast, for the non-helical case as studied in Ref.~\cite{Di:2020ivg} without considering the sphaleron decay, MFs produced during the PT evolves with the scaling law $B_{rms}=B_\star (\xi_M/\xi_\star)^{-(\theta+1)/2}$ for $\theta=1,4$.  
	The intergalactic MF (IGMF) and the intergalactic magnetic helicity can be constrained with gamma-ray and cosmic-ray observations~\cite{AlvesBatista:2018owq,AlvesBatista:2016urk,Tashiro:2013ita,Batista:2021rgm,Vachaspati:2020blt}\footnote{Ref.~\cite{Tashiro:2013ita,AlvesBatista:2016urk,Batista:2021rgm} found the helical MF of the strength $10^{-14}$ G which may be able to set bounds on low scale PT. }. 
	In the near future, the Cherenkov Telescope Array (CTA) observations will provide the sensitivity up to $10^{-13}$ G on the IGMF, which may discriminate the helicity of the MFs from the PT. For illustration, we present the case with $\beta/H_\star=10$. We find that the fractional helicity generated with the sphaleron decay that can seed the cosmic MF is still beyond the sensitivity of CTA. 
	
	Finally, we note that our result can be used to make some predictions for  more general scenarios of PT in the sense that a smaller sphaleron rate would reduce the fractional helicity $\epsilon_M$ further and therefore reduce the MF strength at the correlation length; and as studied in Ref.~\cite{Di:2020ivg}, fewer bubbles being generated during the PT with large inverse duration ( $\beta/H_\star$) will yield a larger bubble wall velocity\footnote{It still hold in the expanding
		Universe as can be found in Ref.\cite{Guo:2020grp}.} with the MF spectrum shifting to larger peak frequency regions slightly that corresponds to a smaller correlation length of the MF.  
	
	\noindent{\it \bfseries Conclusions:}
	\label{sec:conclusions}
	In this Letter, we have performed a simultaneous study of the Chern-Simons number evolution and the MF production during the first-order electroweak PT. 
	The sphaleron decay is observed during the first-order electroweak PT process proceeding with bubble nucleations, collisions, and coalescence process with the generated MFs being partially helical. The change of the Chern-Simons number is observed associated with the sphaleron decay, which provides the $B+L$ violation and the possible explanation of the BAU. 
	Given the evolution of magnetohydrodynamics turbulence after the PT, we obtain MFs at the characteristic correlation length~\cite{Kahniashvili:2012uj,Brandenburg:2017neh} that may seed the observed MFs in galaxy clusters. Our results suggest that for the first-order electroweak PT with nucleation rate being $\beta/H_\star\sim\mathcal{O}( 10)$: 1) the scenario of the MFs generation from the PT without considering sphaleron decay with $\theta=1,4$ have already been excluded by observations of
	Fermi-LAT and VERITAS; 2) the MFs generated by the case with $\epsilon_M\leq 0.2$ and $\theta=1$ can be probed by CTA in the future; and 3) the scenario of MFs generated during the PT with $\beta/H_\star\leq 10$, together with much larger $\epsilon$ and $\theta$ that can seed large scale MFs is still beyond the current lower bound of blazars and future CTA.
	Therefore, the signature observed here can be used to distinguish whether MFs are totally from bubble collisions or partially from sphaleron decay.
	
	We note that the future large lattice sizes with much smaller resolutions are required to precisely fix the power-law indexes of the MF spectrum in the ultra-infrared regions and ultraviolet regions. And to settle down the exact fractional MF helicity, one needs to accurately determinate the sphaleron rate and the change of the Chern-Simons number during the first-order PT which is absent in the literature. For the previous studies of the electroweak sphaleron rate in the SM, we refer to Ref.~\cite{DOnofrio:2014rug} and references therein.
	We left these two issues for future investigations.
	
	\noindent{\it \bfseries Acknowledgements}
	We thank Daniel Cutting, Anders Tranberg, Paul Saffin, Zong-Gang Mou,
	Tanmay Vachaspati, Andres Diaz-Gil, Juan Garcia-Bellido, Margarita Garcia Perez, Antonio Gonzalez-Arroyo, Rafael Alves Batista, Hiroyuki Tashiro, Misao Sasaki, and Kazunori Kohri for helpful communications and discussions.  
	Ligong Bian was supported by the National Natural Science Foundation of China under the grants Nos.12075041, 12047564, and the Fundamental Research Funds for the Central Universities of China (No. 2021CDJQY-011 and No. 2020CDJQY-Z003),  and Chongqing Natural Science Foundation (Grants No.cstc2020jcyj-msxmX0814).
	RGC is supported by the National
	Natural Science Foundation of China Grants No.11690022, No.11821505, No. 11991052, No.11947302, and by the Strategic
	Priority Research Program of the Chinese Academy of Sciences Grant No. XDB23030100 and the Key Research Program
	of Frontier Sciences of CAS.

	\bibliographystyle{apsrev}
	
	\bibliography{GWBPT}
	
\end{document}